\documentstyle[aps,epsf]{revtex}

\newcommand{\bee}{\begin{equation}}
\newcommand{\ee}{\end{equation}}
\newcommand{\bea}{\begin{eqnarray}}
\newcommand{\eea}{\end{eqnarray}}
\newcommand{\R}{\rm I\kern-.2emR}
\newcommand{\C}{\rm \kern.25em\vrule height1.4ex
depth-.12ex width.06em\kern-.31em C}
\newcommand{\N}{{\rm I\kern-.16em N}}
\newcommand{\Z}{{\rm Z\kern-.35em Z}}

\begin{document}                                                                
\draft
\twocolumn[\hsize\textwidth\columnwidth\hsize\csname
@twocolumnfalse\endcsname
\title{Absence of Asymptotic Freedom in Non-Abelian Models
}

\vskip 1.0truecm
\author{Adrian Patrascioiu}
\address
{\it Physics Department, University of Arizona,
 Tucson, AZ 85721, U.S.A.}
\author{Erhard Seiler}
\address{\it Max-Planck-Institut f\"ur Physik
(Werner-Heisenberg-Institut)
F\"ohringer Ring 6, 80805 Munich, Germany}
\date{\today}
\maketitle
\begin{abstract}
The percolation properties of equatorial strips of the two dimensional 
$O(3)$ nonlinear $\sigma$ model are investigated numerically. Convincing 
evidence is found that a sufficently narrow strip does not percolate at 
arbitrarily low temperatures. Rigorous arguments are used to show that 
this result implies both the presence of a massless phase at low 
temperature and lack of asymptotic freedom in the massive continuum limit.
A heuristic estimate of the transition temperature is given which is 
consistent with the numerical data.
\end{abstract}
]
\narrowtext
\vskip2mm

One of the crucial unanswered problems in particle and condensed matter
physics concerns the phase diagram of the two dimensional ($2D$) nonlinear
$\sigma$ models \cite{www}. 
It is widely believed that the nonabelian models with
$N \geq 3$ are in a massive phase for any finite inverse temperature
$\beta$ and that this is intimately related to their perturbative asymptotic
freedom. Over the years we have brought forth many reasons why we think
that these beliefs are unfounded \cite{apz,ff,super},
but the absence of a mathematical
proof combined with ambiguous numerical results left the issues wide open.
In the present letter we would like to provide convincing numerical
evidence that in fact the $2D$ $O(3)$ model  possesses a massless
phase for sufficiently large $\beta$ and give a rigorous proof
that this is incompatible with asymptotic freedom in the massive phase.
We will also give a heuristic explanation of why and where 
the phase transition happens.

The models we are considering consist of unit length spins $s$ 
taking values on the sphere $S^{N-1}$, placed at the sites of a $2D$ 
regular lattice. These spins interact ferromagnetically with their 
nearest neighbors. Let $\langle ij \rangle$ denote a pair of neighboring 
sites. We will consider two types of interactions between neighbouring 
spins:
\begin{itemize}
\item Standard action (s.a.):  $H_{ij}=-s(i)\cdot s(j)$
\item Constrained action (c.a.):  $H_{ij}=-s(i)\cdot s(j)$
for $s(i)\cdot s(j) \geq c$
and
$H_{ij}=\infty$ for $s(i)\cdot s(j)<c$ for some $c\in [-1,1)$.
\end{itemize}

Almost a decade ago we showed \cite{ael} that one can rephrase the 
issue regarding the existence of a soft phase in these models as a 
percolation problem and in fact this is the reason we introduced the c.a.
model (please note that the c.a. model has the same perturbative
expansion as the s.a. model and possesses instantons).
Namely let $\epsilon=\sqrt{2(1-c)}$ and $S_\epsilon$ the
set of sites such that $|s\cdot n|<\epsilon/2$ for some given unit vector
$n$. Our rigorous result was that if on the triangular (T) lattice the
set 
$S_\epsilon$ does not contain a percolating cluster, then the
$O(N)$ model must be massless at that $c$. For the abelian $O(2)$
model we could prove the absence of percolation of this equatorial set
$S_\epsilon$ for $c$ sufficiently large \cite{ael} (modulo certain
technical assumptions which were later eliminated by  M. Aizenman 
\cite{aiz}). For the nonabelian cases we could not give
a rigorous proof. We did however present certain arguments \cite{ap,npb} 
explaining why the percolating scenario seemed unlikely.

In this letter we will give convincing numerical evidence that a
sufficiently narrow equatorial strip
does not percolate for any $c$. We will also show that via a rigorous
inequality derived by us in the past \cite{conf}, the existence
of a finite $\beta_{crt}$ in the s.a. model on the square (S) lattice is
incompatible with the presence of asymptotic freedom in the massive
continuum limit.

For clarity we will review first a few crucial points regarding 
percolation in $2D$:
\newline a) Let $A$ be a subset of the lattice defined by the spin
lying in some subset ${\cal A}\subset S^{N-1}$ and $\tilde A$ its 
complement. Then
with probability 1 $A$ and $\tilde A$ do not percolate at the same time. 
(For this point it is crucial that the lattice is self matching, hence 
our use of the T instead of the S lattice; on the latter an 
ordinarily connected cluster can be stopped by a cluster connected only 
$\star$-wise, i.e. also diagonally.) This fact has been proven rigorously
only for special cases like the $+$ and $-$ clusters of the Ising model,
but 
is believed to hold generally. 
\newline b) If neither $A$ nor its complement $\tilde A$ percolate,
then the expected size of the cluster of $A$ attached to the origin,
denoted by $\langle A\rangle$ of $A$, diverges; the same holds for its
complement $\tilde A$  (Russo's lemma \cite{russo}). (If $\tilde A$ 
percolates, then $\langle A\rangle$ is finite.)

Let then ${\cal P}_\epsilon$ (the union of the polar caps) be the 
complement of ${\cal S}_\epsilon$ (the equatorial strip of width 
$\epsilon$). According to the discussion above, either
$S_\epsilon$ percolates, or $P_\epsilon$ percolates, or neither
$S_\epsilon$ nor $P_\epsilon$ percolates and then both have divergent
mean size (we shall call this third possibility in short {\it formation 
of rings}). Consider a sequence of tori of increasing size $L$ and the
mean cluster size of the set $A$ corresponding to a subset 
${\cal A}\subset S^{N-1}$ of positive measure. If $A$ percolates
$\langle A\rangle=O(L^2)$, if its complement percolates 
$\langle A\rangle$ will approach a finite nonzero value, and if $A$ forms
rings $\langle A\rangle=O(L^{2-\eta})$ for some $\eta>0$. Therefore,
if we define the ratio
\bee
r=\langle P_\epsilon\rangle/\langle S_\epsilon\rangle     ,
\ee
for $L\to\infty$ it should either go to 0, or to $\infty$ or to some 
finite nonzero value depending on which one of the three possibilities is
realized; the latter possibility assumes that $\eta$ is the same for 
$P_\epsilon$ and $S_\epsilon$, as indicated by our numerics
and consistent with scaling.

\begin{figure}[htb]
\centerline{\epsfxsize=6.0cm\epsfbox{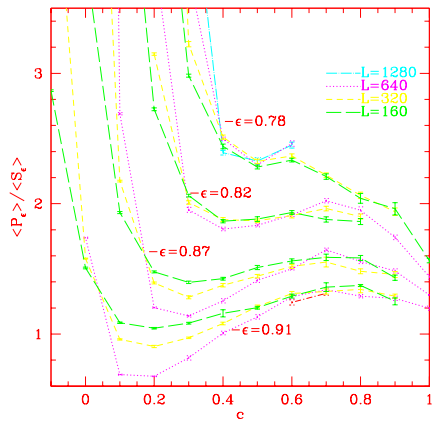}}
\caption{Ratio $\langle{P_\epsilon\rangle/\langle S_\epsilon\rangle}$
for various $\epsilon$ values versus $c$}
\label{tr}
\end{figure}

In Fig.\ref{tr} we show the numerical value of the ratio $r$ as function
of $c$ for $\beta=0$ for four values of $\epsilon$ for the c.a. model on 
a T lattice. 

The results were obtained from a Monte Carlo (MC) investigation using an 
$O(3)$ version of the Swendsen-Wang cluster algorithm
and consist of a minimum of 20,000 lattice configurations used
for taking measurements. For each value of $\epsilon$ we studied $L=160$,
320 and 640 (for $\epsilon=.78$ we also studied $L=1280$). Three distinct
regimes are manifest for each of the four values of $\epsilon$ 
investigated: 
\begin{itemize}
\item For small $c$, $r$ is increasing with $L$, presumably diverging
to $\infty$ (region 1).
\item For intermediate $c$, $r$ is decreasing with $L$, presumably
converging to 0 (region 2).
\item For $c$ sufficiently large depending upon $\epsilon$, $r$ becomes
independent of $L$, just as it does at the crossings from region 1 into 2.
\end{itemize}

\begin{figure}[htb]
\centerline{\epsfxsize=6.0cm\epsfbox{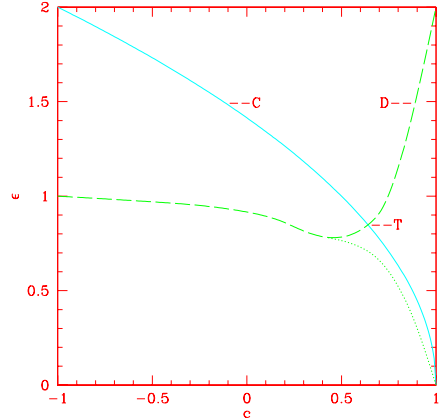}}
\caption{Phase diagram of the $O(3)$ model on the T lattice}
\label{phase}
\end{figure}

Consequently for these values of $\epsilon$ for small $c$
$P_\epsilon$ percolates, for intermediate $c$ $S_\epsilon$
percolates and for sufficiently large $c$ both $P_\epsilon$ and
$S_\epsilon$ form rings. In Fig.\ref{phase} we present the phase
diagram of the c.a. model on the T lattice for $\beta=0$.
The dashed line $D$ represents the minimal equatorial width above which
$S_\epsilon$ percolates. For $c=-1$ (no
constraint)
the model reduces to independent site percolation, for which the 
percolation threshold is known rigorously to be $\epsilon=1$. The rest
of the diagram represents qualitatively the results of our investigation
of the ratio $r$, such as those shown in Fig.\ref{tr}. Two features
of this diagram are worth emphasizing:
\newline 1. An equatorial strip of width less than approximately $\epsilon=.76$
{\it never} percolates.
\newline 2. For approximately $c>0.4$ a new phase opens up in which both 
$P_\epsilon$ and $S_\epsilon$ form rings
(the dotted line separates it from percolation of $P_\epsilon$).

For clarity let us briefly review the argument indicating that this phase
diagram is incompatible with the existence of a massive phase for
arbitrarily large $c$. Choosing an arbitrary unit vector $n$, one
introduces Ising variables $\sigma=\pm 1$. The s.a. Hamiltonian
becomes:
\bee
H_{ij}=\sigma_i\sigma_j |s_\|(i) s_\|(j)|+s_\perp(i)\cdot s_\perp(j)
\ee
where $s_\|(i)=s(i)\cdot n$ and $s_\perp(i)=(s(i)\times n)\times n$. One 
thus obtains an induced Ising model for which the Fortuin-Kastleyn (FK) 
representation \cite{fk} is applicable. In this representation the Ising 
system is mapped into a bond percolation problem: between any like 
neighboring Ising spins a bond is placed with probability 
$p=1-\exp(-2\beta s_\|(i)s_\|(j))$. For the c.a. model a bond is also
placed if after flipping one of the two neighboring Ising spins the
constraint $s(i)\cdot s(j) \geq c$ is violated. The FK representation relates
the mean cluster size of the site clusters joined by occupied bonds 
(FK-clusters) to the Ising magnetic susceptibility. In a massive phase 
the latter must remain finite. Hence, if the FK-clusters have divergent 
mean size, the original $O(3)$ ferromagnet must be massless (the Ising 
variables $\sigma$ are local functions of the originally spin variables
$s$).

Now notice that by construction for the c.a. model the FK-clusters with
say $\sigma=+1$ must contain all sites with $s(i)\cdot n>\sqrt{(1-c)/2}$.
Therefore the model must be massless if clusters obeying this condition
have divergent mean size. But the polar set $P_\epsilon$ consists
of two disjoint components $P^+_\epsilon$ (north) and $P^-_\epsilon$ 
(south). For $c>(1-\epsilon^2)/2$ there are no clusters containing
both elements of $P^+_\epsilon$ and $P^-_\epsilon$. Hence if for such 
values of $c$ clusters of $P_\epsilon$ form rings, so do clusters of
$P^+_\epsilon$ separately and hence the $O(3)$ model must be massless. 
The curve $C$ given by $c=(1-\epsilon^2)/2$ is the solid line in
Fig.\ref{phase}. The point $T$ at the intersection of the curves $D$ and 
$C$ gives an upper bound for $c_{crt}$, the value of $c$ above which the
c.a. model must be massless.

\begin{figure}[htb]
\centerline{\epsfxsize=6.0cm\epsfbox{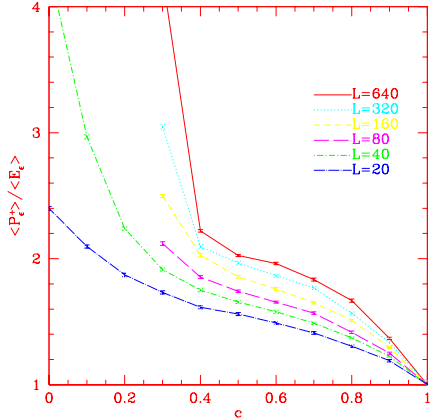}}
\caption{The ratio of the mean cluster size of 
a polar cap of height .75 to that of an equatorial
strip of the same height}
\label{rings}
\end{figure}

To verify the phase diagram in Fig.\ref{phase} we also measured (at 
$\beta=0$) the ratio of the mean cluster size of the set $P^+_{\epsilon'}$
with $\epsilon'=.5$ to that of the set $S_\epsilon$ with $\epsilon=0.75$,
for which the equatorial strip never percolates (Fig.\ref{rings})
($\epsilon$ and $\epsilon'$ ar chosen such that the two sets have equal
density). The data indicate that for some intermediate values of $c$
$P^+_{\epsilon'}$ forms rings while $S_\epsilon$ has finite mean size; this
region terminates around $c=0.4$, where also $S_\epsilon$ starts forming
rings. 
The larger average cluster size of the polar cap compared to the strip 
of the same area is probably due to the fact that the strip has a
larger boundary than the polar cap. 
This is in agreement with a general conjecture stated in
\cite{ap}, namely that for $c$ sufficiently large, if two sets have equal area
but different perimeters, the one with the smaller perimeter will
have larger average cluster size. For the case at hand, this is apparently
true for all values of $c$.

\begin{figure}[htb]
\centerline{\epsfxsize=6.0cm\epsfbox{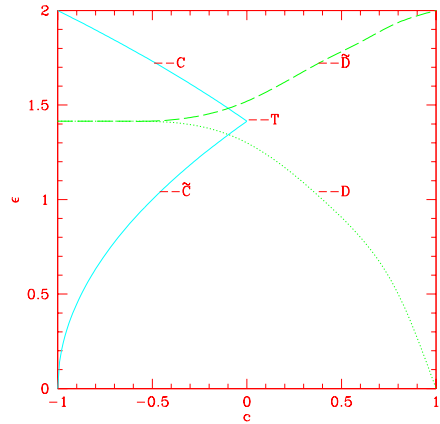}}
\caption{The phase diagram for the $O(2)$ model on the T lattice}
\label{phaseo2}
\end{figure}

The general belief, which we criticized in ref.\cite{ff},is that there is
a fundamental difference between abelian and nonabelian models. To test 
this belief we studied the ratio $r$ for the c.a. $O(2)$ model on the 
T lattice. The phase diagram is shown in Fig.\ref{phaseo2}. 
Since in the $O(2)$ model the set ${\cal P}_\epsilon$ can also be 
regarded as a set ${\cal S}_{\tilde \epsilon}$ where 
$\tilde \epsilon=\sqrt{(4-\epsilon^2)}$, certain features of that diagram 
follow from rigorous arguments. For instance it is clear that in the
c.a. model there exist two curves $C$ and $\tilde C$ 
and in the region to their right the model must be 
massless \cite{ael,aiz}. The precise location of the curves
$D$ (or $\tilde D$) must be determined numerically, 
something which we did not do. We did verify though that the ring 
formation region begins around $c=-0.5$.

In our opinion the arguments and numerical evidence provided so far give
strong indications that the c.a. models on the T lattice possess
a massless phase. 
Universality would suggest that a similar situation must exist for the 
s.a. models on the T and S lattices. To test universality we measured
on the S lattice the renormalized coupling
both on thermodynamic lattices in the massive phase and in finite volume
in the presumed critical regime (as in \cite{kp}). Our data for the c.a.
model on the S lattice only determine an interval (about $.5$ to $.7$) in
which the
massless phase of the model sets in; we tried to see if we could get a
similar $L$ dependence for the renormalized coupling in the s.a. 
model at a suitable $\beta$ as for $c=.61$ in the c.a. model at
$\beta=0$.
This seems to be indeed the case for $\beta$ roughly $3.4$. We went only
up to $L=640$, hence this equivalence between $c$ and $\beta$ should be
considered only as a rough approximation, but there seems to be no doubt
that the two models have the same continuum limit.
 
It is intersting to note that there is a heuristic explanation for both
the existence of a massless phase in the s.a. $O(3)$ model and for the 
value of $\beta_{\it crt}$. Indeed it is known rigorously that in $2D$
a continuous symmetry cannot be broken at any finite $\beta$. 
In a previuos paper \cite{super} we 
showed that the dominant configurations at large $\beta$ are not 
instantons but superinstantons (s.i.). In principle both instantons and s.i.
could enforce the $O(3)$ symmetry. In a box of diameter $R$ the 
former have a minimal energy $E_{\it inst}=4\pi$ \cite{bp} while the
latter $E_{s.i.}=\delta^2\pi/\ln R$, where $\delta$ is the angle by which
the spin has rotated over the distance $R$. Now suppose that 
$\beta_{\it crt}$ is sufficiently large for classical configurations to 
be dominant. Then let us choose $\delta=2\pi$ (restoration of symmetry) 
and ask how large must $R$ be so that the superinstanton configuration 
has the same energy as one instanton. One finds $\ln R=\pi^2$. But in the
Gaussian approximation

\bee
\langle s(0)\cdot s(x)\rangle\approx 1-{1\over \beta\pi}\ln x
\ee

Thus restoration of symmetry occurs for $\ln x\approx\pi\beta$. This
simpleminded argument suggests that for $\beta\geq\pi$ instantons become
less important than s.i.. Now in a gas of s.i. the image of any small
patch of the sphere forms rings, hence the system is massless. While this
is not a quantitative argument, we believe it captures qualitatively what
happens: a transition from localized defects (instantons) to s.i..

Next let us discuss the connection between a finite $\beta_{crt}$
and the absence of asymptotic freedom. It follows from our earlier work
concerning the conformal properties of the critical $O(2)$
model \cite{conf}. We refer the reader for details to that paper and give
only an outline of the argument. The s.a. lattice $O(N)$ model possesses
a conserved isospin current. This currrent can be decomposed into a
transverse and longitudinal part. Let $F^T(p)$ and $F^L(p)$ denote the 
thermodynamic values of the 2-point functions of the transverse and
longitudinal parts at momentum $p$, respectively. Using reflection 
positivity and a Ward identity we proved that in the massive continuum 
limit the following inequalities must hold for any $p\neq 0$:
\bee
    0\leq F^T(p)\leq F^T(0)=F^L(0)\leq F^L(p)=2\beta E/N
\ee
Here $E$ is the expectation value of the energy 
$$E=\langle s(i)\cdot s(j)\rangle$$ at inverse temperature $\beta$. Since
$E \leq 1$ it follws that if $\beta_{\it crt}<\infty$ $F^T(0)-F^T(p)$ 
cannot diverge for $p\to\infty$ as required by perturbative asymptotic 
freedom. In fact, for $\beta_{crt}=3.4$ (which is a plausible guess)
$F^T(p)$ must be less than 2.27, in violation of the form factor 
computation giving $F^T(0)-F^T(\infty)>3.651$ \cite{bn}.

Since the implications of our result, that for the c.a. model a 
suffciently narrow equatorial strip never percolates, are so dramatic, 
the reader may wonder how credible are the numerics. The usual suspect, 
the random number generator, should not be important since
precision is not the issue here. The only debatable point is whether our
results represent the true thermodynamic behaviour for $L\to\infty$ or 
are merely small volume artefacts. While we cannot rule out rigorously
that scenario, certain features of the data make it highly implausible:
\begin{itemize}
\item Small volume effects should set in gradually, while the data in
Fig.\ref{tr} indicate a rather abrupt change from a region where $r$ is
decreasing with $L$ to one where $r$ is essentially independent of $L$.
\item
For $c\to 1$ at fixed $L$, $r$ must approach the `geometric' value 
$r=2/\epsilon-1$. As can be seen, in all the cases studied, throughout 
the `ring' region $r$ is clearly larger than this value,
while it should go to 0 if $S_\epsilon$ percolated.
\item In Fig.\ref{rings} there is 
no indication of the ratio going to 0 for increasing $L$. Moreover the
dramatic change in slope around $c=.4$ indicates that the polar cap
$P_{\epsilon'}$ starts forming rings at a smaller value of $c$ than 
the equatorial strip $S_\epsilon$.
\end{itemize}
Thus we doubt very much that the effects we are seeing represent small
volume artefacts. Moreover, if $s_z$ remained massive at low temperature 
and in fact an arbitrarily narrow equatorial strip percolated, one would 
have to explain away our old paradox \cite{ap,npb}: if such a
narrow strip percolated, an even larger strip would percolate and on it 
one would have an induced $O(2)$ model in its massless phase, in 
contradiction to the Mermin-Wagner theorem.

Consequently we are confident that the phase diagram in Fig.\ref{phase}
represents the truth, that a soft phase exists both in the s.a. and the
c.a. model and that the massive continuum limit of the $O(3)$ model is
not asymptotically free. In a previous paper \cite{super} we have already
shown that in nonabelian models even at short distances perturbation theory 
produces ambiguous answers. The present result sharpens that result by
eliminating the possibility of asymptotic freedom.

Acknowledgement: AP is grateful to the Humboldt Foundation for a
Senior US Scientist Award and to the Werner-Heisenberg-Institut for its
hospitality.

\end{document}